\def\br{{\bf r}}
\def\bR{{\bf R}}
\def\bk{{\bf k}}
\def\bkp{{\bk '}}

\def\bd{{\bf d}}
\def\bB{{\bf b}}
\def\w0{\omega_0}
\def\k0{k_0}
\def\wk{\omega_k}

\def\ekj{\hat{e}_{\bk j}}

\def\ekjp{\hat{e}_{\bk 'j'}}

\def\bkj{\hat{b}_{\bk j}}

\def\bkjp{\hat{b}_{\bk 'j'}}

\def\akj{a_{\bk j}}
\def\akjd{a_{\bk j}^\dagger}

\def\kvac{\mid \widetilde{0}\rangle}
\def\bvac{\langle \widetilde{0}\mid}

\documentclass[twocolumn,showpacs]{revtex4}

\begin{document}

\title{Vacuum local and global electromagnetic self-energies for a point-like and an extended field source}
\author{Roberto Passante, Lucia Rizzuto, and Salvatore Spagnolo}
\affiliation{Dipartimento di Fisica e Chimica dell'Universit\`{a} degli Studi di Palermo and CNISM, Via Archirafi 36, I-90123 Palermo, Italy}

\email{roberto.passante@unipa.it}

\pacs{12.20.Ds, 03.70.+k}

\begin{abstract}
We consider the electric and magnetic energy densities (or equivalently field fluctuations) in the space around a point-like field source in its ground state, after having subtracted the spatially uniform zero-point energy terms, and discuss the problem of their singular behavior at the source's position. We show that the assumption of a point-like source leads, for a simple Hamiltonian model of the interaction of the source with the electromagnetic radiation field, to a divergence of the renormalized electric and magnetic energy density at the position of the source. We analyze in detail the mathematical structure of such singularity in terms of a delta function and its derivatives.
We also show that an appropriate consideration of these singular terms solves an apparent inconsistency between the total field energy and the space integral of its density. Thus the finite field energy stored in these singular terms gives an important contribution to the self-energy of the source.
We then consider the case of an extended source, smeared out over a finite volume and described by an appropriate form factor. We show that in this case all divergences in local quantities such as the electric and the magnetic energy density, as well as any inconsistency between global and space-integrated local self-energies, disappear.
\end{abstract}

\maketitle

\section{\label{sec:1}Introduction}

Vacuum fluctuations and the existence of the zero-point energy of the electromagnetic field are a remarkable prediction of quantum electrodynamics \cite{Milonni}. They have been extensively investigated in the literature,
especially in relation to Casimir and Casimir-Polder forces, which are long-range electromagnetic
interactions between neutral macroscopic objects (metals or dielectrics), between atoms and surfaces or among neutral atoms or molecules \cite{C48,CP48,Lifshits56,MPRSV08}. The Casimir effect, in particular, is related to the modification of the zero-point energy as a consequence of a change of the boundary conditions on the electromagnetic field.

Despite its oddity, the idea of a force generated by a change of vacuum fluctuations in the presence of boundary conditions has stimulated great interest in the literature, both from theoretical and experimental
point of view. These purely quantum effects have now been experimentally confirmed with remarkable accuracy \cite{SBCSH93,Lamoreaux97,MR98,BKMM09} and many efficient methods have been proposed to calculate the Casimir force between bodies for real materials \cite{BKMM09} and nontrivial geometries \cite{MW02,KK06,KK08,MNLR08,EGJK07,REGJK09,MREGJK11}, even if some controversies are still present in the literature.

A controversial issue concerns with the appearance of surface divergences (and their cut-off dependence) in the calculation of field energy densities, in the presence of metallic boundary conditions \cite{DC79,MCW06,FS98,Pfenning00, Milton11a}.
The physical origin of these divergences has been recently questioned in the literature and the possibility of removing them through a suitable  regularization procedure has been discussed
in the case of a scalar field \cite{FS98,Milton11a}. Generally speaking, the appearance of surface divergences in the field energy density is not surprising: boundary conditions are in fact a convenient oversimplification
of the interaction between matter and fluctuating fields. An ideal boundary constraints all field modes at any wavelength, and this gives rise to singular energy densities on the boundary.
For example, it is well-known that, in the case of a perfectly reflecting plate, renormalized electric and magnetic energy densities (i.e. after subtraction of the homogeneous energy density existing also in absence of the boundary)
diverge at the vacuum-conductor interface. These quantities can be made finite by introducing an appropriate exponential upper cut-off in the frequency integrations \cite{Pfenning00,BP12}. It has been also shown that they disappear in the case of a boundary with a fluctuating position \cite{FS98}; on the other hand, it was shown that, in the cases of a dielectric half space \cite{SF02} and a dielectric sphere \cite{BK02}, dispersion does not remove such divergences.

Recently,  the electromagnetic field fluctuations near a dielectric-vacuum interface have been investigated \cite{HL98,Pfenning00}. The structure of the surface divergences in the limit of a perfect conductor has been discussed in detail \cite{BP12}. Also, local energy densities and surface divergences have been explored near fluctuating boundaries \cite{FS98} or in the vicinity of a {\em soft } wall modeled
by a potential which is monomial in the distance from the wall \cite{Milton11}, or in the presence of background fields \cite{GJKQSW04,BK08}.

In spite of the increasing interest on the subject, many topics are not yet well understood. For example,
the presence of   divergences in the field energy density
presents serious problems when the coupling to gravity is considered, because considerable gravitational effects should be observed \cite{Milton11a,MNS11}. In fact, even if considering real surfaces might eliminate the divergences associated to idealized boundaries, energy densities at surfaces and their effects could be large and should be carefully considered \cite{EFM12,EFKKLM08}. It should be stressed that electromagnetic energy densities can be experimentally investigated through the Casimir-Polder interaction energy with appropriate electrically and magnetically polarizable bodies \cite{PP87}.

All the considerations above make relevant investigating the structure and physical origin of the divergences of the vacuum energy densities in the presence of boundaries or polarizable bodies. In \cite{BP12} we have investigated the origin and structure of the surface divergence at the interface between an ideal conductor and the vacuum space, exploiting an appropriate limit procedure from a dielectric to a metal; this has also allowed us to show that surface divergences in the electromagnetic energy density are essential in order to have consistency between the space integral of the field energy density and the global field self-energy.

It is thus natural to investigate in more detail also the case of vacuum electromagnetic energies near a field source such as an atom or a polarizable body, and this will be the main subject of this paper.

In this paper we study the electromagnetic energy densities (or, equivalently, the field fluctuations) surrounding a dressed point-like and a dressed  extended source of the electromagnetic field, whose properties are given by its polarizability and form factor; in particular, we shall concentrate on their behavior at the source's position. Compared to previous works on the subject mentioned above, in the present case we have a source of the electromagnetic field in place of a boundary condition.
Some aspects of the vacuum fluctuations of the electromagnetic field in presence of a point-like source have been already explored in the literature (see for example \cite{CPPP95} and references therein), as well as their connection with Casimir-Polder interactions between atoms and/or macroscopic bodies \cite{PP87}, but their singular behavior was not considered. Surface divergences of the energy density have been recently considered in the case of a thin spherical shell \cite{CPMW06} and a cylindrical shell \cite{CPMK07}.

We first consider a simple Hamiltonian model for the interaction of a polarizable body with the quantum electromagnetic field \cite{CP69}, often used for the calculation of retarded Casimir-Polder forces. In this model the source properties are given by its static polarizability and the effective interaction is quadratic in the coupling constant. This model correctly reproduces the retarded renormalized electric and magnetic energy densities surrounding a point-like ground-state atom, proportional to $1/r^7$ ($r$ being the distance from the atom), after subtraction of the vacuum (zero-point) energy terms that we are not calculating in this paper.  They clearly diverge at $\br=0$, as well as their sum and their integral over all space. On the other hand, the renormalized total field energy (i.e. the expectation value of the field Hamiltonian, after subtraction of the zero-point terms) can be shown to be zero in this model (see Sec. \ref{sec:2}). This shows an inconsistency between local and global self-energies, analogous to the case of the ideal metallic boundary condition discussed in \cite{BP12}. In that case it was shown that only a careful consideration of the surface divergence at the metal-vacuum interface permits to restore a full consistency between the two approaches. This gives a strong indication that, also in the case of a point-like field source, extra singular terms at the atomic position should be present.
It has been argued, in a different situation, that a singular behavior may arise from the oversimplified assumption of point-like source and dipole approximation \cite{CPP88,CPP88a}. In some sense, point-like sources are conceptually similar to ideal boundaries, so singular energy densities at the position of the source could be expected.

The field quantities considered can be made finite by a suitable regularization procedure involving
the use of an appropriate upper-frequency cut-off in the calculations.
We introduce an exponential cut-off function in the frequency integrals, and let the cut-off frequency go to infinity only after having obtained the distance-dependence of the energy density, spending specific attention to its value at $\br =0$. We thus show, for the point-like source, the presence of extra singular terms localized at $\br =0$, in addition to the well-known $1/r^7$ term. Also, this procedure allows us to prove that there is not any inconsistency between global and local field self-energies, provided the energy-density divergences localized at the source's position are properly included. Also, our results clearly show that the finite energy stored in the singularity gives an essential contribution to the field self-energy of the source.

We then consider the case of an extended  field source of finite dimension interacting with the electromagnetic field in the vacuum state. In this case we use a more general Hamiltonian model, with a frequency-dependent polarizability.
The source is modeled as a collection of elementary neutral sources, smeared out over a finite volume
with a density described by a function $\rho(\br)$, whose Fourier transform gives the form factor of the source.
This form factor, which appears in the matter-field interaction Hamiltonian, plays the role of a regularization factor in the source-field interaction and provides a finite length-scale that cuts off the contribution of high frequency modes.
This ensures that the electric and magnetic energy densities are finite everywhere, thus eliminating the problem of the divergence present in the point-like source case. Global and local self-energies are fully consistent in this case, even for a vanishing size of the source. The case of a point-like source can then be obtained by an appropriate limit procedure.

This paper is organized as follows.
In Sec. \ref{sec:2} we consider a point-like source interacting with the quantum electromagnetic field, in its dressed ground-state. We assume the source located at $\br=0$ and investigate the local and global (i.e. integrated over all space) properties of the
electric and magnetic energy densities in the vacuum space surrounding the source, after that the vacuum (zero-point) energy has been subtracted.
We show that the assumption of a point-like source leads to a divergence in the electromagnetic energy density, and obtain their explicit mathematical expressions.
We show that a singular behavior at $\br=0$ is at the origin of the apparent discrepancy between the total electromagnetic self-energy, obtained as the expectation value of the field Hamiltonian or as a space integral of the field energy density.
We find that such inconsistency is completely removed if the singularity at the source position $\br=0$ is correctly taken into account,
by virtue of a subtle cancelation between quantities diverging at $\br=0$.
In Sec. \ref{sec:3} we consider the case of an extended source with a finite size and analyze in detail the behavior of the field energy density in the space around the source.
We discuss how the proposed model of extended source eliminates divergences and singularities in the electric and magnetic energy densities, and also solves the mentioned inconsistency in the calculation of the total electromagnetic energy for any nonvanishing size of the source. Sec. \ref{sec:4} is finally devoted to some concluding remarks.

\section{\label{sec:2} Electric and magnetic energy densities around a point-like source}
In order to discuss the problem of the divergences of the renormalized electromagnetic energy densities, we first consider the case of  a point-like source in its ground-state interacting with the electromagnetic field in the
vacuum state. We suppose the source located at $\br=0$ and describe its interaction with the field by
a simple effective Hamiltonian model, frequently used for the calculation of retarded interatomic Casimir-Polder forces \cite{CP69}
\begin{eqnarray}
\label{eq 2.1}
H = H_A + H_F-\frac 1 2 \alpha \bd^2(0)
\end{eqnarray}
(we are working in the Coulomb gauge). $H_A$ is the Hamiltonian of the source (atom or polarizable body) and $H_F$ is the electromagnetic field Hamiltonian;
$\bd_{}(\br)$ is the transverse displacement field that, outside the source, in this coupling scheme coincides with the total (transverse plus longitudinal) electric field, and $\alpha$ the static polarizability of the source. Introducing the usual bosonic annihilation and
creation operators $\akj$ and $\akjd$, the Hamiltonian of the radiation field assumes the well-known expression
\begin{equation}
\label{eq 2.4}
H_F = \frac{1}{8\pi} \int (\bd^2(\br)+\bB^2(\br)) d^3\br = \sum_{\bk j} \hbar \wk a^{\dagger}_{\bk j}a_{\bk j},
\end{equation}
where the zero-point energy has been subtracted in the last equality, with
\begin{eqnarray}
\label{eq 2.2}
\bd_{}(\br) = i \sum_{\bk j}\left(\frac{2\pi\hbar\wk}{V}\right)^{1/2} \! \! \ekj \left( \akj e^{i \bk\cdot \br} -  \akjd e^{-i \bk\cdot \br}\right) \\
\label{eq 2.3}
\bB(\br)= i \sum_{\bk j}\left(\frac{2\pi\hbar\wk}{V}\right)^{1/2} \! \! \bkj \left( \akj e^{i \bk\cdot \br} - \akjd e^{-i \bk\cdot \br}\right)
\end{eqnarray}
where $\ekj$ are the polarization vectors assumed real and
$\bkj= \ekj\times \hat{\bk}$.

In this model for the source-field coupling, the source's internal degrees of freedom are \emph{frozen}, similarly to the macroscopic case of a perfectly conducting boundary or the case of a static source in quantum field theory \cite{HT62}.

This model allows us to obtain analytical expressions for the electric and magnetic energy density and shows the existence of extra singular terms located at the source's position giving a finite contribution to the field energy, without burdening the results with unessential details on the structure of the source.

We are interested in investigating local and global properties of the virtual electromagnetic field surrounding the point-like source in its dressed ground state, specifically its energy density and its total energy.
In particular, we shall calculate the quantum average of the electric and magnetic energy densities on the dressed ground state, that is the lowest-energy eigenstate of the total Hamiltonian (\ref{eq 2.1}).
Using Rayleigh-Schr\"{o}dinger perturbation theory,
the dressed ground-state is easily obtained at first order in the polarizability $\alpha$,
\begin{eqnarray}
 \label{eq 2.5}
 \kvac&=&\mid g, 0_{\bk j} \rangle -\frac{\pi \alpha}{V}\sum_{\bk j}\sum_{\bkp j'} \frac{(kk')^{1/2}}{k+k'} \hat{e}_{\bk j}\cdot\hat{e}_{\bkp j'} \nonumber \\
 &\times& \mid g, 1_{\bk j}1_{\bkp j'}\rangle ,
\end{eqnarray}
where $g$ denotes the atomic ground state.

We now evaluate the expectation values of the squared electric and magnetic fields on the dressed ground-state (\ref{eq 2.5}).  After straightforward algebraic calculations we get, at order $\alpha$,
\begin{eqnarray}
\label{eq 2.6}
& &\bvac\mathcal{H}_{el}(\br) \kvac = \frac{1}{8\pi}  \bvac \bd^2(\br) \kvac\nonumber\\
&=& \frac{\hbar c \alpha}{4\pi} \left(\frac{\pi}{V}\right)^2\sum_{\bk j}\sum_{\bkp j'}(\ekj\cdot\ekjp)^2\frac{kk'}{k+k'}\nonumber\\
&\times& \! \! e^{i(\bk+\bkp)\cdot \br}+ c.c. ,\\
\label{eq 2.7}
& &\bvac\mathcal{H}_{m}(\br) \kvac=\frac{1}{8\pi}  \bvac\bB^2 (\br) \kvac\nonumber\\
&=& \frac{\hbar c \alpha}{4\pi} \left(\frac{\pi}{V}\right)^2\sum_{\bk j}\sum_{\bkp j'}(\ekj\cdot\ekjp)(\bkj\cdot \bkjp)\frac{kk'}{k+k'}\nonumber\\
&\times& \! \! e^{i(\bk+\bkp)\cdot \br}+ c.c. ,
 \end{eqnarray}
where we have subtracted the spatially uniform zero-point contributions, thus obtaining the \emph{renormalized} energy densities (from now onwards, we shall only discuss quantities obtained after this subtraction).

In the continuum limit ($V \to \infty$), after sums over polarizations and angular integrations, we get
\begin{eqnarray}
\label{eq 2.8}
& &\bvac\mathcal{H}_{el}(\br) \kvac=\frac{\alpha\hbar c}{4\pi^3}\int_{0}^{\infty} \! \! dk \int_0^{\infty} \! \! dk' \nonumber\\
& &\times Q_E(k,k',r) \frac{k^3 k'^3}{k+k'} \\
\label{eq 2.9}
& &\bvac\mathcal{H}_{m}(\br) \kvac=-\frac{\alpha\hbar c}{4\pi^3}\int_{0}^{\infty}  \! \! dk \int_0^{\infty} \! \! dk' \nonumber\\
& &\times Q_M(k,k',r) \frac{k^3 k'^3}{k+k'} ,
\end{eqnarray}
where $Q_E(k,k',r)$ and $Q_M(k,k',r)$ are given in terms of spherical Bessel functions as
\begin{eqnarray}
& &Q_E(k,k',r)=j_0(kr)j_0(k'r) - j_0(kr)\frac{j_1(k'r)}{k'r} \nonumber\\
& &-\frac{j_1(kr)}{kr}j_0(k'r)+\frac{3}{k k' r^2} j_1(kr)j_1(k'r),
\label{eq 2.10}\\
& &Q_M(k,k',r)=2 j_1(kr)j_1(k'r).
\label{eq 2.10a}
\end{eqnarray}
Using the relation
\begin{eqnarray}
\label{eq 2.13}
\frac{1}{k+k'}=\int_0^{\infty} \! \! e^{-(k+k')\eta} d\eta
\end{eqnarray}
to decouple $k$ and $k'$ integrations in (\ref{eq 2.8}) and (\ref{eq 2.9}),
after integration over $k$, $k'$ we get
\begin{eqnarray}
\label{eq 2.11}
\bvac\mathcal{H}_{el}(\br) \kvac &=& \frac{4\alpha\hbar c}{\pi^3}
\int_0^{\infty} \! \! d\eta\frac{3 r^4-2r^2\eta^2+3\eta^4}{(r^2+\eta^2)^6},\\
\label{eq 2.12}
\bvac\mathcal{H}_{m}(\br) \kvac &=&-\frac{4\alpha\hbar c}{\pi^3}
\int_0^{\infty}  \! \! d\eta\frac{8 r^2\eta^2}{(r^2+\eta^2)^6}.
\end{eqnarray}
Integration on $\eta$ finally gives the well-known results (valid for $r \neq 0$)
\begin{eqnarray}
\label{eq 2.14}
& &\bvac\mathcal{H}_{el}(\br) \kvac = \frac{23}{(4\pi)^2}\frac{\alpha\hbar c}{r^7} ,\\
\label{eq 2.15}
& &\bvac \mathcal{H}_{m}(\br) \kvac= - \frac{7}{(4\pi)^2}\frac{\alpha\hbar c}{r^7}
\end{eqnarray}
The expressions above give the (local) electric and magnetic energy density surrounding the point-like source. These energy densities around the field source are proportional to the two-body Casimir-Polder interaction energy with an appropriate electrically or magnetically polarizable test body placed at a certain distance from the source, as shown in Refs. \cite{PP87,CPPP95}.
They are well defined everywhere, except at the origin $\br=0$, where they diverge. Before exploring in more detail this singular behavior and showing the existence of additional singular terms at $\br =0$,
it is worth to consider a global field quantity such as its total energy on the dressed ground state. Using (\ref{eq 2.4}) and (\ref{eq 2.5}), we immediately obtain
\begin{equation}
\label{eq 2.16}
\bvac  H_F \kvac
= \bvac\sum_{\bk j} \hbar \wk a^{\dagger}_{\bk j}a_{\bk j} \kvac= 0 .
\end{equation}
On the other hand, equations (\ref{eq 2.14}) and (\ref{eq 2.15}) imply
\begin{equation}
\label{eq 2.17}
\bvac \left( \mathcal{H}_{el}(\br) +  \mathcal{H}_{m}(\br) \right) \kvac
= \frac{1}{\pi^2}\frac{\alpha\hbar c}{r^7} ,
\end{equation}
which diverges when integrated over all space because of the behavior of (\ref{eq 2.14}) and (\ref{eq 2.15}) for $\br =0$.
This is at variance with the result (\ref{eq 2.16}), because
$H_F = \int d^3r ( \mathcal{H}_{el}(\br) +  \mathcal{H}_{m}(\br))$.
There is thus a discrepancy between the value of the total electromagnetic field energy, calculated as the expectation value (\ref{eq 2.16}) of the field Hamiltonian $H_F$ or as a space integral of the electromagnetic energy density (\ref{eq 2.17}).
It appears as if the average field energy could not be obtained from the space integral of (\ref{eq 2.17}); also, the latter is divergent at the source's position, as well as its space integral.
The mathematical origin of this difficulty
is that the energy densities as given in (\ref{eq 2.14}) and (\ref{eq 2.15}) have a singularity at $\br=0$; we shall show in the next part of this Section that additional singular terms are present and that their existence solves this (apparent) paradox.
This singular behavior indeed prevents from exchanging the $r$ and $\eta$ integrations in the calculation of the total electric and magnetic energy  \cite{CPP88}.
To strengthen this observation, we note that if the integration over $r$ is first performed, and then that over $\eta$ (that is the opposite order of the $r$ integral of (\ref{eq 2.17})),
the space integral of the total (electric plus magnetic) energy density correctly vanishes, as expected from (\ref{eq 2.16}).
In fact, from eqs. (\ref{eq 2.8}) and (\ref{eq 2.9}), after integrations on $k$ and $k'$, or from (\ref{eq 2.11}) and (\ref{eq 2.12}), we have
\begin{eqnarray}
\label{eq 2.19}
& &\int d^3r \bvac \mathcal{H}_{el}(\br) \kvac = \frac{16\hbar c \alpha}{\pi^2}
\int_{\eta_m}^{\infty}d\eta\int_0^{\infty}dr r^2 \nonumber\\
& & \times\frac{3 r^4-2r^2\eta^2+3\eta^4}{(r^2+\eta^2)^6}= \frac{\hbar c \alpha}{\pi^2}
 \int_{\eta_m}^{\infty}d\eta \frac{3\pi}{4\eta^5},\\
\label{eq 2.20}
& &\int d^3 r \bvac \mathcal{H}_{m}(\br) \kvac= - \frac{16\hbar c \alpha}{\pi^2}
\int_{\eta_m}^{\infty}d\eta\int_0^{\infty}dr r^2 \nonumber\\
& & \times \frac{8 r^2\eta^2}{(r^2+\eta^2)^6} = -\frac{\hbar c \alpha}{\pi^2}
\int_{\eta_m}^{\infty}d\eta \frac{3\pi}{4\eta^5},
\end{eqnarray}
where a regularization factor $\eta_m \to 0$ has been introduced. From (\ref{eq 2.19}) and (\ref{eq 2.20}) we obtain
\begin{eqnarray}
\label{eq 2.21}
\int \! \! d^3r\bvac \left( \mathcal{H}_{el}(\br)+ \mathcal{H}_{m}(\br) \right) \kvac = 0,
\end{eqnarray}
because the electric and magnetic contributions cancel each other, even if they are individually divergent when $\eta_m \to 0$.
We can physically understand this result: the exponential function introduced in (\ref{eq 2.13}) plays the role of a cut-off function  in the frequency integrations in  (\ref{eq 2.8}) and (\ref{eq 2.9}), and,  before the integration over $\eta$, it removes the singular behavior of energy densities at $\br=0$.  Exchanging the order of the $\eta$ and $r$ integrations is somehow equivalent to remove the cut-off, and this leads to the diverging result, similarly to the case of an ideal conducting boundary leading to diverging energy densities at the boundary-vacuum interface \cite{BP12}.

From the previous considerations, we are led to conclude that without an appropriate prescription for dealing with the singularity at the origin, the total field energy cannot be obtained by a space integration of its energy density.
This observation strongly suggests to investigate in more detail the singular behavior of the electromagnetic energy density.

We now show that,
if the singularity at $\br=0$ is correctly evaluated and taken into account, the space integral of
$\bvac \mathcal{H}_{el}(\br)+ \mathcal{H}_{m}(\br)\kvac$ vanishes, consistently with (\ref{eq 2.16}).
We first consider the electric and magnetic energy densities, following a procedure similar to that used in \cite{CPP88} in a different context. It involves the use of an exponential cut-off  in  the integrals (\ref{eq 2.8}) and (\ref{eq 2.9}), and letting the cut-off frequency go to infinity after the frequency integrations (details are given in the Appendix A).  After some algebra we get
\begin{eqnarray}
\label{eq 2.22}
& &\bvac\mathcal{H}_{el}(\br) \kvac = \frac{\hbar c \alpha}{(4\pi)^2}\left\{ \frac{23}{r^7}-\frac{23}{r^6} \delta (r) +\frac{10}{r^5}\delta'(r)\right.\nonumber\\
& &\left.-\frac{7}{3r^4}\delta''(r)
+\frac{1}{3r^3}\delta'''(r)+\frac{1}{15r^2}\delta^{(iv)}(r)\right\} , \\
\label{eq 2.23}
& &\bvac \mathcal{H}_{m}(\br) \kvac = -\frac{\hbar c \alpha}{(4\pi)^2}\left\{ \frac{7}{r^7}-\frac{7}{r^6} \delta (r) +\frac{2}{r^5}\delta'(r)\right.\nonumber\\
& &\left.+\frac{1}{3r^4}\delta''(r)
-\frac{1}{3r^3}\delta'''(r)-\frac{1}{15r^2}\delta^{(iv)}(r)\right\} ,
\end{eqnarray}
where the superscript to the delta function indicates the order of its derivative with respect to $r$. The first term on the right-hand side of (\ref{eq 2.22}) and (\ref{eq 2.23}) coincides with (\ref{eq 2.14}) and (\ref{eq 2.15}), respectively. The other terms, involving a delta function and its higher-order derivatives, take into account the singularity of the electric and magnetic energy densities at $\br=0$. This singularity was not included in the calculation yielding
(\ref{eq 2.14}) and (\ref{eq 2.15}). Our procedure outlined in the Appendix A, i.e. introducing first an exponential cut-off function and taking the limit of the cut-off frequency to infinity only after the frequency integrals, has thus allowed us to obtain the correct and complete expression of the energy-density singularity at $\br =0$.

We can now easily evaluate the total energy of the field. Summing up (\ref{eq 2.22}) and (\ref{eq 2.23}), after integration over all space and using the properties of the distributional derivatives of $\delta-$functions, we finally obtain
\begin{eqnarray}
\label{eq 2.24}
& & \int \! \! d^3r \bvac \left( \mathcal{H}_{el}(\br)+ \mathcal{H}_{mag}(\br)\right)\kvac
= \frac{\hbar c \alpha}{4\pi}\int_0^{\infty} dr \nonumber\\
& & \times \left\{
\frac{16}{r^5}-\left[\frac{16}{3r^4}-\frac{48}{15r^4}+\frac{28}{15r^4}\right]\delta(r)\right\} = 0
 \end{eqnarray}
(cancelation of terms diverging for $r \to 0$ should be noted). The existence of the integral over $r$ in (\ref{eq 2.24}) also shows that the energy density, as obtained from (\ref{eq 2.22}) and (\ref{eq 2.23}), is a mathematically well-defined quantity as a distribution in all points of space. As mentioned, equation (\ref{eq 2.24}) shows that, by a proper accounting of the divergences at the atomic position, the total field self-energy vanishes, consistently with the global self-energy obtained in (\ref{eq 2.16}).  This means that the total (diverging) electromagnetic energy for $r>0$ is exactly canceled by the electromagnetic energy {\em stored} at $\br=0$, as clearly shown by equation (\ref{eq 2.24}).
This important property is not obtained if the integral of the energy density is performed without a careful consideration of the singularity at $\br=0$. It also shows that the renormalized field energy vanishes by virtue of an intriguing cancelation among diverging quantities in its electric and magnetic components.  This fully restores consistency between the expectation value of $H_F$ on the dressed ground-state and the field energy density integrated over all space. Although the total renormalized self-energy vanishes, its individual electric and magnetic components do not.

We wish to stress that a similar discrepancy has been discussed in the case of quantum fields with a boundary condition, for example for the electromagnetic field in presence of a perfectly conducting plate \cite{FS98, BP12}; it has been shown
that the existence of a singular energy density at the boundary is necessary to remove the inconsistency between the total field energy and its integrated energy density   \cite{BP12}.
We finally observe that the exponential cut-off we have introduced in the previous calculations is just a mathematical procedure that allows us to obtain the correct form of the divergences of energy densities through a limit procedure.  In the next Section we shall discuss how
the assumption of a model of extended source may provide a natural physical cutoff frequency, which removes all divergences in the energy densities of the electromagnetic field.

\section{\label{sec:3} Energy densities and fluctuations of the electromagnetic field near an extended source}

We now focus on the second main point of this paper, specifically the local and global electromagnetic self-energy in the case of an extended source.
We consider a neutral source of finite dimension, which we model
as a collection of neutral point-like sources, smeared-out over a finite volume, with density $\rho(\br)$. This introduces a form-factor in the frequency integrations. Specific examples of form factors for transitions between levels of the hydrogen atom can be found in \cite{Moses73}. We assume the function $\rho(\br)$ normalized, so that
\[
\int d^3\br \rho(\br)=1 .
\]
A generalization of Hamiltonian (\ref{eq 2.1}) to the case of an isotropic source of finite dimensions with a frequency-dependent polarizability is
\begin{eqnarray}
\label{eq 3.1}
H &=&  H_A + \frac{1}{8\pi} \int d^3\br (\bd^2(\br)+\bB^2(\br))- \frac 12 \sum_{\bk j}\alpha(k) \nonumber\\
&\times& \! \! \int d^3\br \rho(\br)\bd_{\bk j}(\br) \cdot
\int d^3\br' \rho(\br')\bd (\br') ,
\end{eqnarray}
where $\bd_{\bk j}(\br )$ is the Fourier component of $\bd (\br)$.
We are now considering in (\ref{eq 3.1}) a more general model than in the previous Section, where the source also has a frequency-dependent dynamical polarizability. The effective Hamiltonian
(\ref{eq 3.1}) is a generalization to the extended source of the effective Hamiltonian introduced in \cite{PPT98,PP87a}.
This Hamiltonian reduces to (\ref{eq 2.1}) for a point-like source $\rho(\br)\rightarrow\delta(\br)$ with a frequency-independent polarizability.
The interaction of an elementary source at $\br '$ with the field in (\ref{eq 3.1}) also depends on the presence of the elementary source at $\br ''$.

The density function $\rho(\br)$ appearing  in the interaction Hamiltonian, acts as a regularizing factor of the source-field interaction. We now show that our extended-source model allows to remove all divergences in the renormalized values of electric and magnetic energy densities, yielding also consistency between the integrated averaged energy density and the average of the field Hamiltonian. This model also permits to obtain the case of a point-like source discussed in the previous Section as a limit case.

In order to evaluate the electromagnetic energy density around the dressed source, we follow
the same procedure of the previous Section. We first calculate the dressed ground state of Hamiltonian (\ref{eq 3.1}): straightforward perturbation
theory up to first order in the atomic polarizability yields
\begin{eqnarray}
\label{eq 3.2}
& & \kvac = {\mid g, 0_{ \bk j} \rangle}
- \frac{\pi}{V}\sum_{\bk j}\sum_{\bkp j'} \alpha(k) \frac{(kk')^{1/2}}{k+k'} \hat{e}_{\bk j}\cdot\hat{e}_{\bkp j'}\nonumber\\
&\times& \! \! \int \! \! \! \int \! d^3\br d^3\br'\rho(\br) \rho(\br')
e^{-i(\bk\cdot \br+\bkp\cdot \br')}
\mid g, 1_{ \bk j}1_{\bkp j'}\rangle .
\end{eqnarray}

This expression can be used to evaluate the quantum averages $\langle \mathcal{H}_{el}(\br)\rangle$ and $\langle \mathcal{H}_{m}(\br)\rangle$. At lowest order in $\alpha (k)$ and subtracting the spatially homogeneous zero-point contributions (present also in absence of the source), we get
\begin{eqnarray}
\label{eq 3.3}
& &\bvac \mathcal{H}_{el}(\bR) \kvac_{e.s.} = \frac{\hbar}{8\pi} \left(\frac{\pi}{V} \right)^2 \sum_{\bk j}\sum_{\bkp j'}
\frac{(kk')^{1/2}}{k+k'}\nonumber\\
& & \times (\hat{e}_{\bk j}\cdot\hat{e}_{\bkp j'})^2  (\alpha(k) +\alpha(k'))e^{i(\bk+\bkp)\cdot \bR}\nonumber\\
& &\times\int \! d^3\br \rho(\br) e^{-i\bk\cdot \br}\int \! d^3\br'\rho(\br') e^{-i\bkp\cdot \br'}
+ c.c.
\end{eqnarray}
for the electric energy density, and
\begin{eqnarray}
\label{eq 3.4}
& &\bvac \mathcal{H}_{m}(\bR) \kvac_{e.s.} = \frac{\hbar}{8\pi} \left(\frac{\pi}{V}\right)^2\sum_{\bk j}\sum_{\bkp j'}
\frac{(k k')^{1/2}}{k+k'}\nonumber\\
& & \times(\ekj\cdot\ekjp) (\bkj\cdot\bkjp) (\alpha(k) +\alpha(k'))e^{i(\bk+\bkp)\cdot \bR}\nonumber\\
& & \times \int \! d^3\br \rho(\br) e^{-i\bk\cdot \br}\int \! d^3\br'\rho(\br') e^{-i\bkp\cdot \br'}+ c.c.
\end{eqnarray}
for the magnetic energy density (subscript $e.s.$ indicates ``extended source''). Comparison with the analogous quantities for the point-like source (which can be obtained from (\ref{eq 3.3}) and (\ref{eq 3.4}) with $\rho (\br ) = \delta (\br )$) allows us to write these quantities in the more compact form
\begin{eqnarray}
\label{eq 3.5}
& &\bvac\mathcal{H}_{el}(\bR) \kvac_{e.s.} =\frac{1}{8\pi}\int d^3\br \rho(\br)\int d^3\br'\rho(\br') \nonumber\\
& &\times\left(\bvac \bd(\bR-\br) \cdot \bd(\bR-\br') \kvac_{p.s.}\right) ,\\
\label{eq 3.6}
& &\bvac\mathcal{H}_{m}(\bR) \kvac_{e.s.} =\frac{1}{8\pi}\int d^3\br \rho(\br)\int d^3\br'\rho(\br') \nonumber\\
& &\times\left(\bvac \bB(\bR-\br)\cdot \bB(\bR-\br') \kvac_{p.s.}\right) ,
\end{eqnarray}
where the subscript $p.s.$ indicates ``point-like source''. Eqs. (\ref{eq 3.5}) and (\ref{eq 3.6}) clearly show the presence of {\em interference} between the contributions of the elementary parts of the extended source.

 Assuming $\rho(\br)$ with a spherical symmetry,  the form factor depends only on $k = \mid \bk \mid$,
 \begin{eqnarray}
 \label{eq 3.7}
 \int d^3\br \rho(\br) e^{- i \bk\cdot\br} = \rho(k).
 \end{eqnarray}
 Substituting (\ref{eq 3.7}) in (\ref{eq 3.5}) and (\ref{eq 3.6}),  in the continuum limit we obtain
 \begin{eqnarray}
 \label{eq 3.8}
 & &\bvac \mathcal{H}_{el}(\bR) \kvac_{e.s.} = \frac{\hbar c }{16\pi^3}\int_0^{\infty} \! \! dk \int_0^{\infty}
 \! \! dk' \left( \alpha (k)\right. \nonumber\\
 & & + \left.\alpha (k') \right) \left( \rho(k)\rho(k') Q_E(k,k',R)
 \frac{k^3k'^3}{k+k'} + c.c. \right)
 \end{eqnarray}
 and
 \begin{eqnarray}
 \label{eq 3.9}
 & &\bvac \mathcal{H}_{m}(\bR) \kvac_{e.s.} =- \frac{\hbar c }{16\pi^3}\int_0^{\infty} \! \! dk \int_0^{\infty} \! \! dk'
 \left( \alpha (k) \right. \nonumber\\
 & & + \left. \alpha (k') \right) \left( \rho(k)\rho(k') Q_M(k,k',R)
 \frac{k^3k'^3}{k+k'}+ c.c. \right) .
 \end{eqnarray}
 These expressions show how the renormalized electric and magnetic energy densities explicitly depend on the structure of the source through its form factor $\rho (k)$.
 In the limit $\rho(r)\rightarrow\delta(r)$ and for $\alpha (k)=\alpha$, they reduce
 to the expressions for a static point-like source obtained in the previous Section.

The assumption of an extended source introduces a physical cutoff (at a frequency related to the size of the source), that makes finite in all points of space the electric and magnetic energy density discussed in the previous Section.
In fact, it is easy to see that the functions inside the integrals in (\ref{eq 3.8}) and (\ref{eq 3.9}) behave for large $k$ as $\alpha (k)\rho(k) k$. Thus, for typical values of $\alpha (k)$ and choosing an appropriate form factor decreasing as $1/k^\alpha$ $(\alpha>2)$, the integrals on the right-hand side of (\ref{eq 3.8}) and (\ref{eq 3.9}) converge. Furthermore, because the functions $Q_{E(M)}(k,k',R)$ are continuous everywhere, both the electric and magnetic  energy densities
are well-defined in any point of space. This also makes convergent their integral over all space.
In other words, the singularity present in the electric and magnetic energy density of the point-like source disappears in the present case of extended source.

Finally, we may explicitly calculate the space integral of the electromagnetic energy density. Now the integral over $r$ can be safely exchanged with the integrals over $k$ and $k'$ due to the regularization introduced by the form factor. Being
\begin{equation}
\label{eq:3.9a}
\int \! \! d^3 r \left( Q_E(k,k',r) - Q_M(k,k',r) \right) = 0 ,
\end{equation}
we immediately get
\begin{equation}
\label{eq 3.10}
\int d^3 r \bvac \left( \mathcal{H}_{el}+\mathcal{H}_{m} \right)\kvac_{e.s.} =0 \; ,
\end{equation}
as expected from eq. (\ref{eq 2.16}), which can be proved also in the case here considered.
This result confirms that, because of the assumption of a finite-dimension source, there is not discrepancy between the total electromagnetic energy and the integrated energy density for any nonvanishing size of the source.

\section{\label{sec:4}Conclusions}

In this paper we have analyzed the vacuum electric and magnetic energy densities (or equivalently electric and magnetic field fluctuations) near a field source, for example an electrically polarizable body, using an effective Hamiltonian model describing the source-field interaction. The importance of considering local quantities such as field energy densities is related to many factors, for example their relation to Casimir-Polder forces on polarizable bodies or their relevance as a source term for gravity. We have first concentrated our interest on the structure of the divergences of the energy densities at the source position ($\br =0)$ in the case of a static point-like source. We have found that analytical expressions of the  electric and magnetic energy densities contain terms proportional to the Dirac delta function and its derivatives evaluated at $\br =0$, which contain a finite amount of energy. We have shown that these singular terms in the energy densities are essential in order to have consistency between the renormalized (i.e. after subtraction of terms existing even in the absence of the source) vacuum expectation values of the field Hamiltonian and of the field energy density integrated over all space. Thus they give an essential contribution to the self-energy of the source. We have also considered a model of an extended source of the electromagnetic field, characterized by a frequency-dependent polarizability and smeared out over a finite volume of space, and shown that in this case the renormalized field energy densities are finite in any point of space for any nonvanishing size of the source. Relation of our results with recent works about surface divergences of the field energy density at the interface between a conducting plate and the vacuum has been also discussed in detail.

\begin{acknowledgements}
The authors wish to thank F. Persico, R. Messina and N. Bartolo for interesting discussions on the subject of this paper.
Financial support by the Julian Schwinger Foundation, by Ministero dell'Istruzione, dell'Universit\`{a} e della
Ricerca and by Comitato Regionale di
Ricerche Nucleari e di Struttura della Materia is
gratefully acknowledged. The  authors acknowledge support from the ESF Research Networking Program CASIMIR.
\end{acknowledgements}

\appendix
\section{\label{AppendixA}Calculation of $\langle \mathcal{H}_{el}(r)\rangle$ and $\langle \mathcal{H}_{m}(r)\rangle$}

In this Appendix we outline the procedure that leads to the expressions
(\ref{eq 2.22}) and (\ref{eq 2.23}) giving the energy-density singularity at $\br =0$. For simplicity, we focus on
the magnetic energy density $\langle \mathcal{H}_m(r)\rangle$, given in (\ref{eq 2.9}). We need to calculate the following integral
\begin{equation}
\label{wq A1}
I(r)= \int_0^{\infty}\! \! dk\int_0^{\infty} \! \!dk' j_1(kr)j_1(k'r) \frac{k^3k'^3}{k+k'}.
\end{equation}
In order to evaluate it, we introduce an exponential cut-off so that this integral becomes
\begin{eqnarray}
\label{eq A2}
I(r)&=& \int_0^{\infty} \! \! dk\int_0^{\infty} \! \! dk' j_1(kr)j_1(k'r) \frac{k^3k'^3}{k+k'}e^{-\gamma(k+k')}\nonumber\\
&=&\int_0^{\infty} \! \! d\eta\left\{\int_0^{\infty} \! \! dk k^3j_1(kr)e^{-(\gamma+\eta)k}\right\}^2,
\end{eqnarray}
with $\gamma >0$ and where we have used relation (\ref{eq 2.13}).  This integral can be easily evaluated; after some algebra we obtain
\begin{eqnarray}
\label{eq A3}
&\ &I(r)=64r^2\int_0^{\infty} \! \! d\eta\frac{(\gamma+\eta)^2}{[r^2+(\gamma+\eta)^2]^6}\nonumber\\
&\ &=\frac{32}{5}\frac{r^2\gamma}{(r^2+\gamma^2)^5}-\frac{4}{5}\frac{\gamma}{(r^2+\gamma^2)^4}-\frac{14}{15}\frac{\gamma}{r^2(r^2+\gamma^2)^3}\nonumber\\
&\ &-\frac{7}{6}\frac{\gamma}{r^4(r^2+\gamma^2)^2}-\frac{7}{4}\frac{\gamma}{r^6(r^2+\gamma^2)}\nonumber\\
&\ &+\frac{7}{4}\frac{1}{r^6}\int_\gamma^{\infty}\frac{d\mu}{r^2+\mu^2}
\end{eqnarray}
where $\mu=\gamma+\eta$.  We now consider the limit $\gamma\rightarrow 0$, equivalent to removing the exponential cut-off, \emph{after} the frequency integrals. Using the Lorentzian representation of the Dirac delta function,
it is easy to see that the terms appearing in the expression (\ref{eq A3}) lead to the Dirac delta function and its derivatives when $\gamma \to 0$.
Thus, we obtain
\begin{eqnarray}
I(r)&=& \frac {\pi}8 \left( \frac{7}{r^7}-\frac{7}{r^6} \delta (r)+\frac{2}{r^5}\delta'(r)
+\frac{1}{3r^4}\delta''(r)\right. \nonumber\\
&-& \left. \frac{1}{3r^3}\delta'''(r)-\frac{1}{15r^2}\delta^{(iv)}(r) \right),
\end{eqnarray}
where the superscript to the delta function indicates the order of its derivative with respect to $r$. Substituting in (\ref{eq 2.9}) and
using (\ref{eq 2.10a}), we finally obtain expression (\ref{eq 2.23}). A similar procedure leads to expression (\ref{eq 2.22}) for the electric energy density.

\end{document}